\documentclass{jltp}

\usepackage{graphicx}
\def\beq{\begin{equation}}
\def\eeq{\end{equation}}
\def\He3{$^3$He}
\def\He4{$^4$He}
\def\Tl{$T_{\lambda}$}

\title{The discovery of superfluidity}

\author{S\'ebastien
Balibar\footnote{e-mail: balibar@lps.ens.fr}}

\address{Laboratoire de Physique Statistique de l'Ecole Normale Sup\'erieure\\
associ\'e aux Universit\'es Paris 6 et 7 et au CNRS\\
24 Rue Lhomond, 75231 Paris Cedex 05, France}

\runninghead{S\'ebastien Balibar}{The discovery of superfluidity}

\begin{document}

\maketitle

\begin{abstract}
Superfluidity is a remarkable manifestation of quantum mechanics at
the macroscopic level.  This article describes the history of its
discovery, which took place at a particularly difficult period of the
twentieth century.  A special emphasis is given to the role of J.F.
Allen, D. Misener, P. Kapitza, F. London, L. Tisza and L.D. Landau.
The nature and the importance of their respective contributions are
analyzed and compared.  Of particular interest is the controversy
between Landau on one side, London and Tisza on the other, concerning
the relevance of Bose-Einstein condensation to the whole issue, and
also on the nature of thermal excitations in superfluid helium 4.  In
order to aid my understanding of this period, I have collected several
testimonies which inform us about the work and attitude of these great
scientists.

\end{abstract}

\section{Introduction}
Scientific progress has become a collective process.  No physicist
can ever pretend that he has achieved something, that he had had a
personal idea or made any original discovery independently of his
colleagues.  Recognizing this situation does not mean that it is
impossible to identify the authors of scientific discoveries, but one
should do it carefully.  Instead, there is some tendency to attribute
discoveries to single persons, an attitude which is not fair enough.
Moreover, our prize tradition is certainly very nice, stimulating and
generous, but it has some drawbacks: one is tempted to forget those
among our colleagues who did not win. With these ideas in mind, I
have found particularly interesting to inquire about the history of
the discovery of superfluidity.

As we shall see, I am not saying that Kapitza in 1978 or Landau in
1962 were awarded the Nobel prize for the discovery of superfluidity,
nor criticizing this choice.  In fact, for Kapitza it was ``for his
basic inventions and discoveries in the area of low-temperature
physics'' and for Landau ``for his pioneering theories for condensed
matter, especially liquid helium''.  Furthermore, in the official
presentation speech of Kapitza's prize, it was mentioned that ``The
same discovery was made independently by Allen and Misener''.
However, since superfluidity occupies a large part in the official
presentation of their prizes, and since nobody else was recognized at
that level for the discovery of superfluidity, there is a general
tendency to forget that other great scientists have achieved major
contributions to this discovery.  It is this tendency which I wish to
criticize.  One example is the article {\it Superfluidity} in the
Encyclopaedia Britannica, which starts with the sentence: ``{\it
Superfluidity in helium-4 was discovered in 1938 by the Soviet physicist
Pyotr Leonidovich Kapitsa.}'' Another striking example is the
presentation speech of the 1996 Nobel prize to Lee, Osheroff, and
Richardson, where one reads: ``{\it it was not until the end of the
1930s that Pjotr Kapitsa (Nobel Prize 1978) discovered experimentally
the phenomenon of superfluidity in helium-4}'' (no mention of Allen and
Misener this time).  As we shall see below, Landau also considered
that superfluidity had been discovered by Kapitza only and he must
have had a strong influence on the opinion of his colleagues; for 
example E.M. Lifshitz wrote:``{\it I have been
asked by the editors of Scientific American to give a short survey of
what has been learned about superfluidity, first discovered in 1937 by
Peter L. Kapitza at the Institute for Physical Problems in 
Moscow}''.\cite{Lifshitz} 
As for the attribution of the theoretical understanding of
superfluidity to Landau, the situation is more subtle, especially
since the discovery of superfluidity  in alkali
gases\cite{Cornell,Ketterle} where the existence of Bose-Einstein 
condensation is obvious, but it is somewhat similar.  For example,
R. Donnelly wrote\cite{Donnelly}
: ``{\it Finally, there was no great scientific leader active in
understanding liquid helium in the early days.  When Kapitza and the
great theoretical physicist Landau, followed by physicists such as
Fritz London, Lars Onsager, Richard Feynman and other greats, came on
board, there was a tremendous surge of excitement, which lasted for
many years and helped bring the subject to its present state of
understanding}''.  I wish to explain that the contributions by London
and by Tisza, which were published three years before Landau's, were major
breakthroughs in the understanding of superfluidity. Fortunately, my
opinion seems to be shared by several other
authors.\cite{Leggett,Nozieres-book}

Some aspects of this issue have already been considered by several
authors, especially by R. Donnelly in the article mentioned
above\cite{Donnelly}, by K. Gavroglu in his biography of Fritz
London\cite{Gavroglu}, by A. Griffin at a summer school on
Bose-Einstein condensation\cite{Griffin} and in his study of ``John
C. McLennan and his pioneering research on superfluid
helium''\cite{Griffin05}, and by J. Matricon and G. Waysand in their 
book\cite{Matricon}.  When trying to go deeper into it, I
distinguished three more precise questions :

1- Who made the experimental discovery ?

2- Who has initiated its theoretical understanding ?

3- How did all this happen in a period (the late 1930's and early
1940's) where the world was torn apart by conflicts and wars ?

One usually considers that superfluidity was discovered in December
1937, the submission date of the two articles on the flow of liquid
helium which appeared side by side in \textit{Nature} on January 8,
1938.  On page 74 was the article by P. Kapitza\cite{Kapitza} and on
page 75 the one by J.F. Allen and A.D. Misener\cite{Allen1}.
As we shall see, very important work was also 
done before, especially in Toronto and in Leiden, but it is really the
publication of these two articles which triggered the theoretical work
of London, Tisza, and Landau. The purpose of this
article is to put everyone's work back in its historical and
scientific context, so that the importance of each contribution could
be judged.  It is also to analyze the very interesting controversy
which opposed Landau to London and Tisza about the role of
Bose-Einstein condensation (BEC) in superfluidity and about the nature
of excitations in superfluid \He4.  In order to understand it I have
recently inquired from Tisza himself, from D. Shoenberg, and from
A. Abrikosov whose testimonies are reproduced here.  I am also
grateful to L. Pitaevskii, G. Volovik, A. Griffin, H. Meyer and G.
Gorelik for several fruitful discussions.  I cannot pretend that I
have fully understood the role and the attitude of every actor in the
discovery of superfluidity but I hope that this article will stimulate
further research on this very important event in the history of
twentieth century physics.

\section{Experiments}

The two articles published in \textit{Nature} are respectively
entitled: ``Viscosity of liquid helium below the lambda point'' (page
74), received December 3, 1937, by P. Kapitza (Institute for Physical
Problems, Moscow) and ``Flow of liquid Helium-II'' (page 75), received
on December 22, 1937, by J.F. Allen and A.D. Misener (Royal Society
Mond Laboratory, Cambridge, UK).  Both the expressions ``lambda
point'' and ``helium II'' refer to the work of W. H. Keesom and his
group in Leiden.  \Tl ~is the temperature now known as 2.17~K where
Keesom, Wolfke and Clusius\cite{Keesom:1927,Keesom:1932} discovered
an anomaly in the properties of liquid \He4 : the graph of the
temperature variation of its specific heat has a sharp maximum with
the shape of the Greek letter $\lambda$.  Thanks to a series of
experiments, Willem Keesom had realized that \He4 had two different
liquid states which he called ``Helium I'' above \Tl , and ``Helium
II'' below (for a review, see Keesom's book\cite{Keesom}).  It must
have been rather surprising to find two different liquid states for
liquid helium which is made of simple spherical atoms without chemical
properties.

In 1937, Kapitza tried to understand why, a year earlier in Leiden,
the same Willem Keesom had found with his daughter
Ania\cite{Keesom:1936} that the thermal conductivity of helium II was
anomalously large, a phenomenon which had also been studied by B.V.
Rollin in Oxford\cite{Rollin} and by J.F.~Allen, R.~Peierls, and
M.Z.~Uddin in Cambridge\cite{Allen:1937}.  Kapitza thought that
convection in this liquid could be important if its viscosity was
small and that it could be responsible for the large apparent thermal
conductivity.  He thus tried to measure this viscosity by flowing
liquid helium from a tube through a slit about 0.5 micrometers thick,
between two polished cylinders pressed against each other.
In his article\cite{Kapitza}, Kapitza writes:
 
``\textit{The flow of liquid above the $\lambda$-point could be only
just detected over several minutes, while below the $\lambda$-point
the liquid helium flowed quite easily, and the level in the tube
settled down in a few seconds. From the measurements we can conclude
that the viscosity of helium II is at least 1500 times smaller than
that of helium I at normal pressure.}

\textit{The experiments also showed that in the case of helium II, the
pressure drop across the gap was proportional to the square of the
velocity of flow, which means that the flow must have been turbulent.
If, however, we calculate the viscosity assuming the flow to have been
laminar, we obtain a value of order 10$^{-9}$ cgs, which is evidently 
still only an upper limit to the true value. Using this estimate, the 
Reynolds number, even with such a small gap, comes out higher than
50,000, a value for which turbulence might indeed be expected}''.
 
These two paragraphs are a little difficult to understand.  Kapitza does
not give any value for the flow velocity in the slit, nor for the
height difference which drove the flow.  Since the Reynolds number is
$ R = UL/\nu$ where $U$ is the velocity, $L$ a typical length scale
and $\nu$ the kinematic viscosity ($\nu = \eta / \rho$ where $\eta$ is
the viscosity and $\rho$ the density), I understand that, with $\eta$
= 10$^{-9}$ cgs, $\rho$ = 0.15 g/cm$^3$, and $L = 5\times 10^{-5}$ cm, he
must have measured velocities $U$ up to about 7 cm/s.  As we shall
see, this is comparable to what had been measured by Allen and Misener
and confirmed by later work, although it depends very much on the size
of the flow (see the review in the book by Wilks\cite{Wilks}, p.
391).  As far as I know, Kapitza's square law for the pressure drop
has not been confirmed by any later work, but since he does not give
much quantitative information on his measurements, it is difficult to
appreciate the accuracy at which this square law could fit his data.
Given what is known today, I suppose that he approximated the pressure
dependence of the velocity -- which shows a threshold -- with a square
law.  Anyhow, Kapitza finally proposes that:

``\textit{by analogy with superconductors, \ldots
the helium below the $\lambda$-point enters a special state
which might be called superfluid}''. 

This is a famous sentence where Kapitza introduces the word
``superfluid'' for the first time.  His intuition was quite remarkable
because superfluids and superconductors are indeed analogous
states of matter, but Kapitza wrote this sentence long before the BCS
theory of superconductivity was established, \textit{a fortiori}
before any demonstration of such an analogy.

As for the article by Allen and Misener, it starts with the sentence:

``\textit{A survey of the various properties of liquid helium II 
has prompted us to investigate its viscosity more carefully. One of
us[1] had previously deduced an upper limit of 10$^{-5}$ cgs units
for the viscosity of helium II by measuring the damping of an
oscillating cylinder. We had reached the same conclusion as Kapitza
in the letter above; namely that, due to the high Reynolds number involved, 
the measurements probably represent non-laminar flow}''.

Before arguing on the question of priority between Moscow and
Cambridge, I wish to comment on the note [1].  It refers to the
article by E.F. Burton published in 1935 by Nature\cite{Burton}.  In
this short article, Burton explains that, by measuring the damping of
an oscillating cylinder which was suspended by a thin rod, it was
possible to measure the viscosity of liquid helium.  This method was
later improved by J.G. Dash and R.D. Taylor\cite{Dash} and again by
F.L. Andronikashvili and by J.D.
Reppy\cite{Andronikashvili,Reppy,Meyer-review} for extensive studies
of superfluidity.  He further explains that, below \Tl , the viscosity
drops down by several orders of magnitude.  He finishes with the
sentence :

``\textit{This work was carried out by Messrs.  Wilhelm, Misener and
A.R.Clark}.''

Burton was the head of the Toronto Physics Department where Misener
was a Master's graduate student at that time, and Wilhelm and Clark
were two
technicians in cryogenics.  The details of this work were later
published by Wilhelm, Misener and Clark in the Proceedings of the
Royal Society\cite{Wilhelm} and I am rather surprised that, at that
time, the head of a physics department could publish work by members
of his department without including their names in the list of
authors.  Since the three real authors of the work published without
including Burton as a co-author, one could imagine that there was a
conflict between them but this is probably not true since, two years
later in his Nature article with Allen, Misener referred to Burton
instead of referring to his own article\ldots I have to suppose that
publication policies have evolved a lot since that time.  It remains
clear that, as soon as in 1935, the existence of an anomaly in the
viscous dissipation in helium II had already been demonstrated in
Toronto.  However, in 1935, no one had realized that the hydrodynamics
of helium II was so anomalous that its viscosity could not be measured
with classical methods.

\begin{figure}
\centerline{\includegraphics[height=7cm]{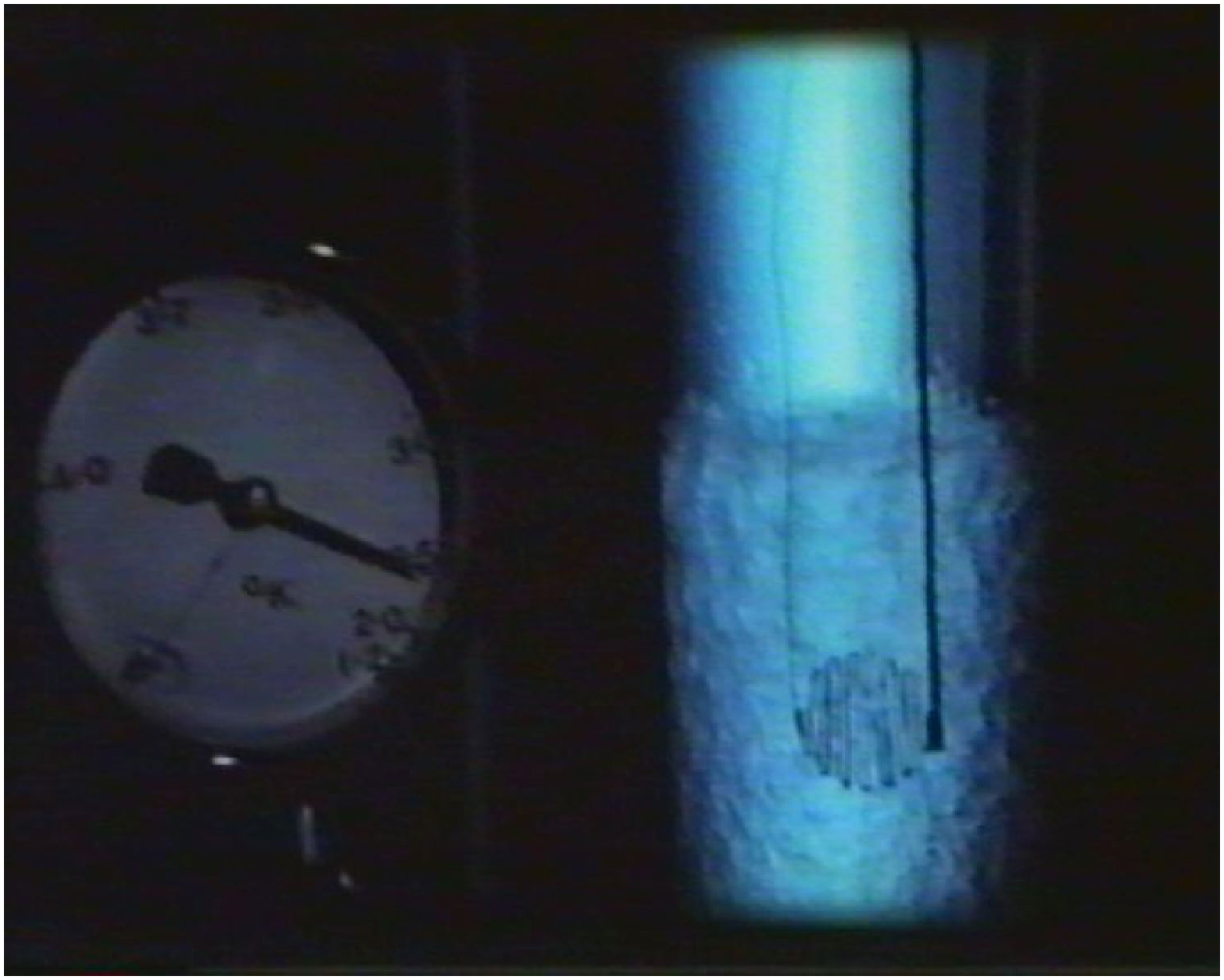}}
\vskip0.5cm
\centerline{\includegraphics[height=7cm]{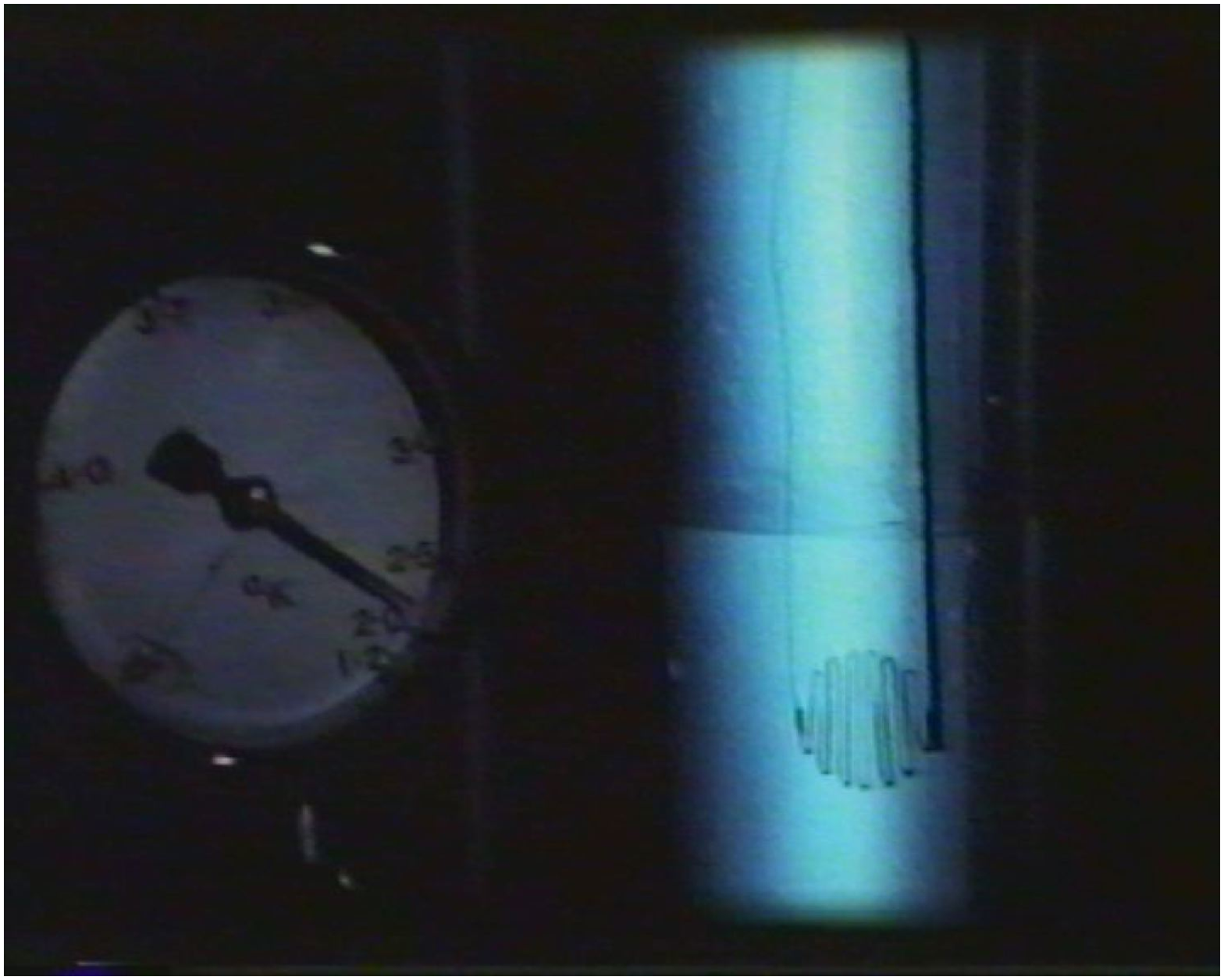}}
\vskip0.5cm
\caption{As shown by these two images from a film by J.F. Allen and
J.M.G.~Armitage, superfluid helium stops boiling below $T_{\lambda}$. This 
is due to its large thermal conductivity. The top image is taken at 
2.4~K as indicated by the needle of the thermometer on the left. The 
bottom image is taken just below the lambda transition.}
\label{fig:boiling}
\end{figure}

At the beginning of his article, Burton also explains that liquid
helium stops boiling when cooled below \Tl .  This phenomenon had been
observed by McLennan three years earlier in Toronto\cite{McLennan}
and it was later attributed to its very large thermal conductivity.
For all physicists working on liquid helium, it remains the
spectacular manifestation of quantum order taking place in this
remarkable liquid (see fig.~\ref{fig:boiling}).

More important is the reference to Kapitza at the beginning of the
article by Allen and Misener. We understand that they had read Kapitza's article
before writing their own, or at least that they had heard of its
content.  Together with the 19 days difference in the submission date,
this has sometimes been taken as a proof that Kapitza had some
priority on Allen and Misener in the experimental discovery of
superfluidity\cite{Andronikashvili}.  However, as we shall see, I do
not agree with such a statement.

The Cambridge article contains a detailed study of the flow through
two different capillaries with sections respectively equal to 6x
10$^{-4}$ and 0.8~mm$^2$.  Measurements are given at two different
temperatures (1.07 and 2.17~K) and at series of ten different
pressures.  Flow velocities range from 0.4 to 14~cm/s.  Their main
findings were that, contrary to Poiseuille's law which describes
laminar situations, the velocity was nearly independent of pressure,
also independent of the capillary section.  The measurements by Allen
and Misener could obviously not be done in 19 days.  I cannot imagine
that they started their study after hearing of Kapitza's article.
If a proof is needed, it is in their notebook which shows that Allen
and Misener had obtained results already on November 24,
1937.\cite{Griffin:2006}

Let us now comment on Kapitza's work.  Kapitza had graduated as an
electrical engineer in Saint Petersburg under the supervision of F.
Ioffe (1918).  In 1921, Ioffe suggested that Kapitza goes to Cambridge
where he could work with Rutherford.  There, he proved to be a
brilliant experimental physicist.  For example, he made the first
detection of the bending of alpha-particle paths in a magnetic field
thanks to a cloud chamber.  Then, he built a pulsed magnetic field
installation and a hydrogen liquefier with his student John Cockcroft.
He was quickly elected Fellow of Trinity College (1925) and Fellow of
the Royal Society (1929) ``{\it a rare distinction for a foreigner,
especially for one who became a Corresponding Member of the Soviet
Academy of Sciences in the same year}'', as explained by David
Shoenberg\cite{Shoenberg:1994}.  Then, Rutherford obtained from the
Royal Society that part of the donation from Ludwig Mond that was used to
build the ``Royal Society Mond Laboratory'' where Kapitza could
develop his low temperature and high magnetic field studies.  In this
laboratory, he constructed a new type of helium liquefier which
produced its first drops of liquid helium on April 19, 1934,
and made such experiments much easier.\cite{Rubinin}  

In the summer of 1934, Kapitza went back to Leningrad.  He had come to
see his mother and to participate in a symposium celebrating the
centenary of Mendeleiev.  However, on September 24, 1934, five months
only after the first operation of his liquefier in Cambridge, he was
not allowed to return to England from the Soviet Union\cite{Rubinin}.
The reasons for this are a little unclear but, according to D.
Shoenberg\cite{Shoenberg:1994}, ``{\it he had sometimes been rather
boastful of his successes in England and gave the impression that his
work could be of immense technological importance if only he were
given the right support.  The authorities, possibly Stalin himself,
took him at his word and told him that he must in the future work for
them, although in fact none of his work was secret and it was
available to everyone}''.  Later, Stalin would need Kapitza for his
nuclear program and conflicts with Beria triggered Kapitza's disgrace.
But in 1934, Kapitza started a fight with Stalin and Molotov to obtain
support for his research. Two years later, the ``Institute for
Physical Problems'' was built for Kapitza in Moscow.  Thanks to the
help of Rutherford, he could also arrange that
part of his equipment be purchased from Cambridge and transferred to
Moscow, so that he could start his research again.

At that stage, one major problem for Kapitza was that Cambridge had
kept his liquefier.  In 1935, liquefiers existed only in Leiden,
Toronto, Cambridge, Oxford, and Kharkov.  But Kapitza also obtained
the right to invite his student David Shoenberg and two technicians,
E.Ya.  Laurmann and H.E. Pearson, in order to build a new helium
liquefier in Moscow.  They made a better liquefier which produced
liquid helium on February 22, 1937.\cite{Rubinin}  Meanwhile,
Cambridge had used Kapitza's rather high salary (800 pounds a year) to
hire two younger scientists, Rudolf Peierls and J.F. Allen who seemed
satisfied with 400 pounds a year each.\cite{Allen:1988}  

John Franck (``Jack'') Allen was born in Winnipeg (Canada) and he had
obtained his PhD on superconductivity in Toronto (1933).  Then, he
tried to join Kapitza in Cambridge but when he arrived in the fall of
1935, Kapitza was already detained in USSR.  In 1936, he
attracted Donald Misener to work towards a PhD degree in Cambridge with him.  We
thus realize that Kapitza was competing with two Canadian physicists
who were using his former liquefier in his former laboratory where
he was still in close contact with other people.  Of course, this
situation was very painful to him (``\textit{ \ldots I often see my
laboratory in my dreams, and painfully want to work\ldots}'' as he wrote
to his wife\cite{Rubinin} in March 1935).  Anyhow, when Kapitza sent
his letter to \textit{Nature}, he wrote in the accompanying letter to
the editor:

``\textit{Dear Gregory,}

\textit{I am sending herewith a short note: `Viscosity
of liquid helium below the $\lambda$-point', which I hope you will
kindly publish in your `letters to the editor'.  I think this is an
important note and I should be glad if you could arrange it to be
published as soon as possible, and with the day of dispatch.  Please
do not bother to send the proofs to me here to Moscow, it takes too
much time.  If necessary please send them either to Prof.  P.A.M.
Dirac, Dr.  J.D. Cockcroft, or to Dr.  W.L. Webster ...  All my good
friends [are] sufficiently competent to make the necessary
corrections.  I hope you will kindly help me in publishing this note
very soon ...  }''

As explained by Allen himself\cite{Allen:1988} and by
Shoenberg\cite{Shoenberg:1994}, it was John Cockcroft\cite{Nobel1951}
who took care of the proof-reading.  He was the new director of the
Mond Laboratory since Kapitza had left.  In December 1937, he showed
Kapitza's letter to Allen and Misener and asked them to write down
their own results as quickly as possible.  He finally asked
\textit{Nature} to publish the two papers side by side.

It is clear to me that the Cambridge work was independent of Kapitza's
work in Moscow.  My main reason is that, as an experimental physicist
in the same field, I know that it is not possible to make all the
measurements which are presented by Allen and Misener in 19 days only.

Now, was Kapitza's work independent of the Cambridge work?  After all,
Kapitza's insistence to be published with a mention of the date of
receipt indicates that he probably knew that his competitors were
working on the same subject.  Furthermore, his letter presents
qualitative ideas which could have been written down quickly.  One
should also note that the ability of helium II to flow through narrow
slits (the existence of ``superleaks'') had been discovered in 1930 by
Willem Keesom.\cite{Keesom1930}  But one does not know if Kapitza was
aware of Keesom's observation.  Could Kapitza have written his letter
{\it after} hearing of the progress made by Allen and Misener in Cambridge?
Nobody having ever mentioned such a possibility, I wish to
consider it carefully.  When I asked David Shoenberg to tell me about
this period, he answered:

``\textit{My memory of the events is not entirely reliable, though I do
remember that I helped translate the Russian version of Kapitza's
letter into English.  Kapitza's letter was sent to Nature with a
request that proofs should go to Cockcroft rather than back to Kapitza
in Moscow.  Also W.L. Webster who had been briefly visiting Kapitza in
Moscow took a copy to show to Cockcroft.  Cockcroft had not, I think,
known of Kapitza's work and showed the note to Jack Allen who had
obtained basically the same result (`superfluidity' below the
$\lambda$ point) and suggested that he writes a brief note (with
Misener) in which he commented on Kapitza's note.  Cockcroft asked
Nature to print the 2 notes side by side but it is quite clear that
Kapitza's note had `official' priority (a) because of the dates of
receipt by Nature and (b) because evidently Allen had seen Kapitza's
note before he wrote his own.  It is a pity Allen never got adequate
recognition of his quite independent discovery of superfluidity -- he
and Kapitza could well have shared a Nobel prize!  I don't think Allen
and Kapitza ever met till much later.  I know that Kapitza was at
first rather cross that his He liquefier was used while he had to wait
a long time before he had liquid helium in Moscow (see one of his
letters to his wife in 1934 or 5). I don't think Kapitza and Allen ever
communicated directly by letter.  I myself know Kapitza was getting
exciting new results while I was in Moscow (as a guest visitor) and
knew that Allen was continuing his work on He (fountain effect etc.
was already published) but I have no memory of discussing the work of
either with the other.  At that time it would have been dangerous to
write to anyone about work going on in Moscow.  I was in Moscow from
September 1937 to September 1938.  I did not travel out of the Soviet Union
at all during that time.}''\cite{Shoenberg}

According to Shoenberg, the work in Moscow was thus independent
from the one in Cambridge because there were no contacts between
Cambridge and Kapitza, but I cannot believe this because letters have
been published by Rubinin\cite{Rubinin} which show the opposite.  For
example, Rutherford sent a letter to Kapitza on October 9,
1937, where he wrote:

``\textit{My dear Kapitza,}

\textit{\ldots Bohr told me about his trip to you [in June 1937], and
I am very interested to hear of the work that you have been able to
accomplish.  No doubt Pearson, when he returns, will be able to give
us the latest information about your big helium liquefier.  The Mond
laboratory is very flourishing, and a large amount of work is in
progress\ldots Some interesting experiments are also in progress on
the extraordinary heat conductivity of helium at low temperatures.
The conductivity is very large for small differences of temperature,
and falls rapidly with the quantity of heat transmitted\ldots}''

This was ten days only before Rutherford died, and Kapitza must have
known this death rather quickly because he sent a letter to John
Cockcroft on the first of November where he wrote:

``\textit{My dear John,}

\textit{It is difficult to believe that there is no more Rutherford\ldots
Things in the lab are not going badly at all. We just started the new 
liquefier and the first time it gave four liters per hour. Now it is
quite certain that Pearson will be free before the new year, I will
not claim his services any more after that\ldots}''

Obviously, there were regular contacts between Kapitza and
his friends in Cambridge.  Furthermore, according to
Rubinin,\cite{Rubinin} Webster had visited Kapitza in Moscow in
September 1937.  As a consequence, it seems to me that Kapitza might
well have known something about Allen's results. However, I am not saying that
Allen and Misener have a priority on Kapitza, in particular because of
a letter from Kapitza to Niels Bohr, dated December 10, 1937, where he
says:

``\textit{Dear Bohr,}

\textit{I had your letter about the death of Rutherford, which apparently
crossed with mine. I had a number of letters from friends, and it is 
indeed wonderful how much the people appreciated Rutherford\ldots
All this time I was very busy working on the viscosity of helium below
the $\lambda$-point. May be you remember what I was telling you during 
your visit here about the idea of the work, the experiments are in
full progress, but the preliminary results are quite interesting. It
appears that really below the $\lambda$-point the viscosity of helium
drops more than a 1000 times\ldots
I made the experiments about 20 times varying the conditions and
looking for some possible errors, but could not find any. I am
sending herewith a copy of my preliminary note to Nature, so if you
will be interested you could glance through it\ldots
Yours very sincerely,}

\textit{P. Kapitza}''

Since Bohr's visit was in June 1937, this letter proves that Kapitza
was at least planning his experiments six months earlier.  It also
shows that Kapitza did much more than a single experiment before
sending his letter to \textit{Nature}. My conclusion on the priority
issue is that there is no priority in either way, the two works are
independent.

Let us finally summarize the content of the four
experimental contributions to the problem of superfluidity.  In 1930
Keesom had observed that helium II was highly fluid and in 1935
Wilhelm, Misener and Clark had measured in Toronto a sharp drop of the
viscosity below the $\lambda$-point.  Then, in December 1937, Kapitza
claimed that the flow of liquid helium II was turbulent and that its
viscosity could not be larger than 10$^{-9}$ cgs units.  As for Allen
and Misener, they presented the results of a series of measurements,
from which they concluded :

``\textit{the observed type of flow\ldots in which the velocity
becomes almost independent of pressure,
most certainly cannot be treated as laminar or even as
ordinary turbulent flow.  Consequently any known formula cannot, from
our data, give a value of the viscosity which would have much
meaning}''.

In my opinion, it is Allen and Misener who discovered that, below \Tl,
the hydrodynamics of helium required a totally new interpretation.  At
that time, everyone else kept considering that liquid helium
was a liquid with a small viscosity.  Here is the real experimental
breakthrough.

It would be very interesting to understand how Kapitza had the
intuition that  helium II had something in common with
superconductors.  The idea that superconductors were quantum systems
which had to be described by a macroscopic wave function had been put
forward by Fritz London and his brother Heinz\cite{London:1935} but,
as we shall see now, London had not yet considered that it could be
the case for liquid helium also.  Furthermore, and as we shall see
when considering Landau's work, this was not the way how one
liked to think about liquid helium in Moscow.

London's new ideas\cite{London:1938} were triggered by the next 
article\cite{AllenJones}  
published by Allen in the same volume 141 of \textit{Nature} on 
February 5, 1938.  Entitled ``New phenomena connected with heat flow in
helium II'', this new letter described the discovery of what is now
known as the fountain effect: together with Misener again for the
experiments, but published with H. Jones, the new young theorist who
replaced Peierls in Cambridge, Allen discovered that, when heat was
applied to liquid helium II on one side of a porous plug, the pressure
increased proportionally to the heat current so that the level of the
free surface went up (it was later realized that the fountain pressure
was in fact proportional to the temperature difference between the two
sides).  A liquid jet could even occur if the pressure
was high enough.  If the liquid had been classical, the vapor pressure
would have been higher on the warm side so that, in order ot maintain
hydrostatic equilibrium in the liquid, its level would have had to go
down.  Allen and Jones explained that the opposite was
observed.  For London, it was no longer possible to doubt that this
liquid had totally anomalous properties for which a radically new
interpretation was needed.  In previous work\cite{London:1936}, Fritz
London had proposed that helium II was more ordered than helium I (its
specific heat decreased sharply below \Tl ) and perhaps some kind of
 crystal with a diamond lattice.  However, on March 5, 1938,
London sent a letter to \textit{Nature} which was published on April
9. There, he explained that liquid helium II was not
crystalline before proposing 
that it was undergoing some kind of Bose-Einstein
condensation at \Tl .\cite{London:1938}

\section{London and Tisza}

In the introduction of his first book\cite{London:book}, 
London, writes:

``{\it In 1924, Einstein developed a very strange concept of a gas of 
identical molecules, which were assumed to be indistinguishable\ldots 
Einstein remarked that this removal of the last vestige of
individuality from the molecules of a species would imply a statistical
preference of the molecules for having the same velocity, even if any 
interaction between them were disregarded, and this preference would
lead, at a well-defined temperature to a kind of change of state of
aggregation; the molecules would `condense' into the lowest quantum 
state, the state of momentum zero. Einstein did not give a very
detailed proof, and his remark received little attention at the time. 
Most people considered it a kind of oddity which had, at best, an
academic interest, for at the extremely low temperatures or high
pressures in question there are no gases, all matter being frozen or
at least condensed by virtue of the intermolecular interaction forces.
In addition, doubt was cast on the mathematical correctness of
Einstein's remark, and hence the matter was disposed of as if there
were no `Bose-Einstein condensation'}.''

On November 29, 1924, Einstein himself had sent a letter to his friend Paul
Ehrenfest in Leiden, where he wrote :

``\textit{From a certain temperature on, the molecules `condense'
without attractive forces, that is, they accumulate at zero velocity. 
The theory is pretty, but is there also some truth in it
?}''.\cite{Pais}

By generalizing the calculation by the young Bengali physicist
Satyendra Nath Bose\cite{Bose} to massive particles, Einstein
had found\cite{Einstein} that, for an ideal gas of Bose particles, a
macroscopic fraction of these particles accumulates in the ground
state below the critical temperature

 \begin{equation}
     T_{BEC}= \left ( \frac{2\pi \hbar^{2}}{1.897 m k_{B}} \right ) 
     n^{2/3}
     \label{eq:BEC}
\end{equation}

At that time, the theory of phase transitions was still in its
infancy, and, in his PhD work\cite{Uhlenbeck:1927}, Uhlenbeck had
argued against the BEC being a true phase transition by saying that it
would not occur in a finite size system.  Uhlenbeck was a graduate
student of Paul Ehrenfest and, apparently, his criticism was generally
accepted, even by Einstein himself.\cite{Griffin} In November 1937,
a conference took place in Amsterdam in honor of van der Waals
(Johannes Diderik van de Waals was born hundred years before, on
November 23, 1837 in Leiden).  Fritz London was there\cite{Gavroglu}
and he must have heard discussions including Ehrenfest and Kramers
about the thermodynamic limit in connection with phase transitions,
also that Uhlenbeck had withdrawn his argument against BEC (see his
publication with Boris Kahn, his student who was later killed by the
Nazis\cite{Kahn:1938}).  This must be what triggered London's
interest in Einstein's forgotten paper on BEC.\cite{Griffin}

In a message which he sent me on the September 4th, 2001, 
Tisza made the following comment on the discovery of superfluidity:

``\textit{The novelty of the effect became strikingly apparent in the
Allen and Jones fountain effect that started London and myself on our 
speculative spree\ldots}''

In his letter to \textit{Nature}\cite{London:1938}, 
Fritz London first recalled that
\He4 ~atoms were Bose particles, then that liquid \He4 ~was a quantum
liquid because the quantum kinetic energy of the atoms was
large, something he had explained in a previous article\cite{London:1936}. 
This large ``zero point energy''
was responsible for the absence of crystallization at 
low pressure, something which had been also noticed by Franz
Simon.\cite{Simon:1934} Then London explained that, although BEC had
``\textit{rather got the reputation of having only a purely imaginary
existence\ldots it actually represents a discontinuity in the
temperature derivative of the specific heat}'', meaning that it was a
phase transition of third order (according to the classification by
Ehrenfest).  Then he calculated the transition temperature $T_{BEC}$
at which an ideal Bose gas with the same density as liquid \He4 ~would
condense in Einstein's sense and he found 3.1~K, a value close to
$T_{\lambda}$.  He further noticed that the singularity in the
specific heat of helium resembled the cusp predicted for BEC. 
He then concluded that, ``\textit{Though the
$\lambda$-point resembles rather a phase transition of second order,
it seems difficult not to imagine a connexion with the condensation of
the Bose-Einstein statistics.  The experimental values of the
temperature of the $\lambda$-point and of its entropy seem to be in
favor of this conception.}'' Keeping this modest attitude, he also estimated
that his model, ``\textit{ which is so far from reality that it
simplifies liquid helium to an ideal gas}'', was a rough approximation
which could not give quantitative agreement with experimental
measurements.  To a modern eye, everything looks right in this
letter to \textit{Nature}. Shortly afterwards, he expanded his
letter into a longer article published the same year.\cite{London:1938b}
London's new ideas created considerable
interest\cite{Gavroglu,Griffin}, in particular from Laszlo Tisza.

Fritz London was born in Breslau (now Wroclaw in Poland) in 1900 and
he had started studies in philosophy before switching to
physics.\cite{Gavroglu} He was educated at the universities of Bonn,
Frankfurt, G\"ottingen and Munich where he graduated in 1921.
Together with Walter Heitler in Zurich, he had devised the first
quantum mechanical treatment of the hydrogen molecule in 1927.  He
then joined Schroedinger in Berlin but in 1933, when the Nazis took
power, he escaped to Oxford where Lindemann found support for him till
1936.  Then, he was quite happy to find a position at the Institut
Henri Poincar\'e in Paris where he was attracted by a group of
intellectuals linked to the ``Front populaire'' (the coalition of
political parties from the French left), namely Paul Langevin, Jean
Perrin, Fr\'ed\'eric Joliot and Edmond Bauer.

Laszlo Tisza had arrived in Paris in 1937 for similar reasons.  He was
born in 1907 and he had studied in Budapest before attending Max
Born's course in G\"ottingen.  Later, he worked in Leipzig under
Heisenberg and wrote his first paper with Edward Teller, just before
being arrested by the Hungarian fascist government under the accusation
of being a communist.\cite{Teller} In 1935, he was liberated and
Teller strongly recommended him to his friend Lev Landau in Kharkov.
There, Tisza entered as number 5 the famous school of theoretical
physics which Landau had founded.  But in March 1937, both Landau and
Tisza left Kharkov.  At this time, Tisza must have tried to protect
himself from anti-Semitism, just like London.  In September 1937, Paul
Langevin and Edmond Bauer offered him a position at the Coll\`ege de
France in Paris.  This is where he met Fritz London; the Coll\`ege de
France is about 300 meters from the Institut Henri Poincar\'e.

Laszlo Tisza told me\cite{Tisza:2001} that they liked discussing
physics together during long walks.  On one such occasion, London
explained his ideas about BEC to him and he had soon 
the intuition that, if BEC took place, there should be two independent
velocity fields in liquid helium.  One part would have zero viscosity
and zero entropy; the other part would be viscous and would carry
entropy; the proportion of each fluid would be related to temperature.
He sent this as a short note to \textit{Nature} on April 16th, 1938,
which introduced for the first time what is now known as the ``{\it two
fluid model}''.\cite{Tisza:1938a}  He announced there more detailed publications which
were presented in French by Paul Langevin at the Acad\'emie des
Sciences on November 14th, 1938, and indeed published in its 
Comptes-Rendus.\cite{Tisza:1938b}

On the basis of his model, Tisza solved the apparent contradiction
between different types of measurement of the viscosity of helium II:
in the Toronto experiment (Wilhelm, Misener, and Clark, 1935), the
damping of the oscillations of the cylinder was related to the
viscosity of the whole fluid while in a flow through a thin capillary
(Allen and Misener, 1937) or through a thin slit (Kapitza, 1937) only
the non-viscous component of the fluid could flow.  He further
explained in this \textit{Nature} note that the independent motion of
the two fluids allowed one to understand the fountain effect.  He
eventually predicted an inverse phenomenon, namely that ``\textit{a
temperature gradient should arise during the flow of helium II through
a thin capillary}''.  The latter was to be named the
``thermomechanical effect'' by Fritz London\cite{London:1938b} and
his brother Heinz\cite{H.London:1938}; evidence for its existence was
found by Mendelssohn and Daunt in Oxford\cite{Mendelssohn} and
further studied by  Kapitza\cite{Kapitza:1941} in 1941.  In the
following articles to the Comptes-Rendus\cite{Tisza:1938b}, Tisza
predicted that, in helium II, heat should propagate as ``{\it temperature
waves}'', another revolutionary idea.  In July 1938, Tisza
``\textit{presented this prediction at a small low temperature meeting
in London\ldots and offered it to make or break [his]
theory}''.\cite{Tisza:2000} His temperature waves were later renamed
``second sound'' by Landau, discovered by Peshkov in 1946 and were indeed
taken as a crucial test of his theory (see below).

At least qualitatively, the 1938 papers by London and Tisza could
explain all the experimental observations which had been already made
at that time, namely the flow and heat conduction experiments, the heat
capacity measurements, also the fast motion of films adsorbed on a
wall by Rollin\cite{Rollin:1936}, confirmed by Daunt and
Mendelssohn.\cite{Daunt}  But still, when London first heard about
Tisza's two fluid model, he could not believe that, in a liquid which
was pure and simple, there could be two independent velocity
fields;\cite{Tisza:1991} 
this was indeed quite a revolutionary idea.  Later, Tisza wrote a more
elaborate version of his theory, which he submitted as two
articles\cite{Tisza:1940} to
the Journal de Physique et du Radium on October 23, 1939, but he could
not see them printed till the end of the war.  In June 1940, part of
Langevin's laboratory was evacuated to Toulouse, in the south part of
France which was not yet occupied by the Nazi army.  In another
e-mail (March 17, 2005), Laszlo Tisza told me that :

``\textit{Jacqueline Hadamard, the daughter of the mathematician
Jacques Hadamard, was a member of the lab and she offered to me and my
wife a ride to Toulouse.  M. and Mme Hadamard traveled with their
other daughter, but I had the privilege to travel as a virtual member
of the Hadamard family.  Just before leaving we had signed up for an
American visa at the Budapest consulate without any definite plans for
using it.  By a fortunate coincidence the Clipper connection between
Lisbon and New York started in the summer 1940 and suddenly we
received air mail letters from friends and relatives in Cambridge in
two days!  We must have notified the Marseille consulate of our
address and sometimes in October we got a telegram that our visa was
authorized.  After finishing all paper work we left Marseille early
February 1941 for Madrid and Lisbon.  Mid-March we sailed on a
Portuguese boat to New York and joined friends and relatives in
Cambridge.  In a few months in September I was appointed instructor at
MIT, to become eventually professor\ldots}''.

As for Fritz London, Frederic Joliot offered him a position of
``Directeur de recherches'' at the Institut Henri Poincar\'e in
November 1938.  He was strongly attached to France where his wife
Edith had entered a group of painters led by Andr\'e
Lhote.\cite{Gavroglu} But he found it wiser to accept an offer from
Paul Gross, the head of the Chemistry Department at Duke University.
He could escape just in time from France, in September 1939, on the
boat ``Ile de France'' to New York, fortunately not on the ``New
Amsterdam'' which was destroyed by a submarine on September 3, three
days after the beginning of the war.\cite{Meyer} In October 1939, Fritz
London was teaching at Duke as a professor of theoretical chemistry.
Nearly at the same time, Landau was coming out of Stalin's jails.

\section{Landau}

Lev Davidovitch Landau was born in Baku on January 22, 1908.  He
graduated from the Physics Department of Leningrad in 1927, at the age
of 19!  He then traveled thanks to a Rockefeller fellowship to
Germany, Switzerland, England and Copenhagen where he worked with
Niels Bohr.  From 1932 to 1937, he was the head of a theory group in
Kharkov.  There, Alexander S. Kompaneets, Evgueny M. Lifshitz,
Alexander I. Akhiezer, Isaak Ya.  Pomeranchuk, and Laszlo Tisza formed
the first core of Landau's famous school; to enter it,
they had to pass the ``{\it Teorminimum exam}''.  At the
same time Landau was also teaching in Moscow and Kapitza invited him
to come in his new Institute in 1937.  However, in March 1938, Landau
was arrested by the NKVD (later called KGB).\cite{Pitaevskii,Gorelik}
He had been accused of being one of the authors of a leaflet
criticizing the Soviet regime.\cite{Gorelik}

Kapitza had already written some letters to Stalin in order to obtain
the scientific equipment he needed for his research in Moscow.  After
Landau's arrest, Kapitza started another fight to liberate him and
eventually sent a letter to Molotov on April 6, 1939,
where, as published in English by P.E. Rubinin,\cite{Rubinin} he wrote :

``\textit{In my recent studies on liquid helium close to the absolute
zero, I have succeeded in discovering a number of new phenomena\ldots
I am planning to publish part of this work\ldots but to do this I need
theoretical help.  In the Soviet Union it is Landau who has the most
perfect command of the theoretical field I need, but unfortunately he
has been in custody for a whole year.  All this time I have been
hoping that he would be released because, frankly speaking, I am
unable to believe that he is a state criminal\ldots It is true that he
has a very sharp tongue, the misuse of which together with his
intelligence has won him many enemies\ldots but I have never noticed
any sign of dishonest behavior\ldots the Soviet Union and worldwide
has been deprived of Landau's brain for a whole year.  Landau is in
poor health and it will be a great shame for the Soviet people if he
is allowed to perish for nothing\ldots }''

Three weeks later, Kapitza was summoned to the NKVD headquarters where
he tried to defend Landau as much as he could in a discussion where he
was asked ``\textit{Do you understand for whom you are pleading?
He's a most dangerous criminal, a spy who confessed to
everything\ldots}''.\cite{Zolitov}  Around 4 o'clock in the morning,
it was said to him: ``\textit{All right, Kapitza, if you pledge your
word for Landau, then give us a written guarantee.  If anything
happens, you will be held responsible}''.  Kapitza wrote a letter to
Beria on April 26, and Landau returned to the Institute on April
28, 1939.  The NKVD decision said:

``\textit{Landau Lev Davydovitch, born in 1908 in Baku, prior to arrest 
professor of physics, non-Party member, and citizen of the USSR, has
been convincingly exposed as a member of anti-soviet group, guilty of 
sabotage and of attempt to publish and disseminate an anti-soviet
leaflet. However, taking into account that (1) Landau LD is a major
specialist in the field of theoretical physics and may be useful in
the future of the Soviet Science; (2) Academician Kapitza PL has
consented to pledge his word for Landau LD; (3) acting on orders of
the People's Commissar\ldots 1st rank Comrade LP Beria to release
Landau in the trust of Academician Kapitza; we hereby order that
detainee Landau LD be discharged from custody, the investigation
discontinued, and case files sent to archive\ldots captain of State
security Vizel}''.\cite{Rubinin}

This allowed Landau to survive and to come back to work. On June 23,
1941, Kapitza\cite{Kapitza:1941} and Landau\cite{Landau:1941a} sent
two letters together for publication in the Physical Review.  They
were published next to each other and Landau's letter announced a more
elaborate paper to be published in the Journal of Physics of the
USSR.\cite{Landau:1941b} 

The two 1941 articles by Landau start with nearly the same sentence:
``\textit{It is well known that liquid helium at temperatures below
the $\lambda$-point possesses a number of peculiar properties, the
most important of which is superfluidity discovered by P.L. Kapitza.}''.
For Landau, superfluidity had thus been discovered by the man who had 
saved his life -- P.L. Kapitza -- and only by him. Landau continues with :

``\textit{L. Tisza[2] suggested that helium II should be considered as a
degenerate ideal Bose gas\ldots This point of view, however, cannot be
considered as satisfactory\ldots nothing would prevent atoms in a
normal state from colliding with excited atoms, i.e. when moving
through the liquid they would experience a friction and there would
be no superfluidity at all. In this way the explanation advanced by
Tisza not only has no foundation in his suggestions but is in direct
contradiction to them}''.\cite{Landau:1941b}

Landau \textit{never} cited Fritz London. Here as everywhere he
attributes to Tisza instead of F. London the proposal that
superfluidity is a consequence of Bose-Einstein condensation. Why is
it that Landau never believed in the relevance of BEC ? This is a
major and quite interesting question. Moreover, why Landau needed to
be so abrupt in his criticism of his former postdoc Tisza? This is a
related question which is no less interesting in my opinion.

After the above introduction, Landau's article starts with a first
section entitled ``The quantization of the motion of liquids''.
Everybody considers what follows as a brilliant breakthrough in the
theory of quantum liquids. He quantizes the hydrodynamics of quantum 
liquids and arrives to the statement ``\textit{Every weakly excited state can 
be considered as an aggregate of single `elementary excitations'} ''
which he divides in two different categories: sound quanta which he
calls ``{\it phonons}'' and which have the linear dispersion relation
\beq
\epsilon = cp
\eeq
and elementary vortices which his friend I.E. Tamm suggested be called
``{\it rotons}'' and for which he proposes the relation
\beq
\epsilon = \Delta + \frac{p^2}{2\mu}
\label{eq:rotons1}
\eeq

In the above equations, $\epsilon$ is an energy, $p$ a momentum, $c$
the sound velocity, $\mu$ an effective mass and $\Delta$ the minimum
energy of rotons (later called the ``{\it roton gap}''). 

Six years later,\cite{Landau:1947}
Landau modified the roton spectrum into

\beq
\epsilon = \hbar \omega = \Delta + \frac{(p-p_{0})^2}{2\mu} \: ,
\eeq
and included them as part of the phonon spectrum.
But already in 1941, Landau could calculate
the specific heat of liquid helium and
obtained a good fit of experimental measurements by W.H. and 
A.P.~Keesom\cite{Keesom:1935} if $\Delta \approx$ 8 to 9~K and $\mu
\approx$ 7 to 8 times the mass of helium atoms.  In his 1941 article, 
Landau then claims
that, for a superfluid flowing at a velocity $V$ at zero temperature,
dissipation can only result from the emission of either phonons or
rotons, so that, from the conservation of energy and momentum in this
process, dissipation is only possible if

\beq
V > V_{cp} = c \:\: {\rm (phonons)\:\: or}\:\: V > V_{cr} = \sqrt{\frac{2\Delta}{\mu}}\: 
{\rm \:\:(rotons)}\: .
\label{eq:critvel}
\eeq

Landau has thus introduced a possible explanation why helium II 
flows at a velocity which is
is independent of pressure or capillary section: his ``critical
velocity'' $V_{c}$ is a property of the helium itself. However, he
also notices that the value he predicts for $V_{c}$ is much larger than
observed in experiments and ``\textit{[left] aside the question as to
whether
superfluidity disappears at smaller velocity for another reason}''.

In the next section he calculates the properties of superfluid helium
at finite temperature.  For this he introduces a two fluid model : he
distinguishes a ``{\it normal component}'' with density $\rho_{n}$, which is
made of phonons and rotons, from a ``{\it superfluid component}'' with
density $\rho_{s} = \rho - \rho_{n}$ ($\rho$ is the total density of
the liquid).  The superfluid component carries no entropy and moves
without dissipation while the normal one is viscous and carries a
non-zero entropy.  The ratio $\rho_{s} / \rho_{n}$ depends on
temperature since, at $T$ = 0, $\rho_{s}$ = $\rho$ and $\rho_{n}$ = 0,
while, at $T$ = $T_{\lambda}$, $\rho_{n}$ = $\rho$.  Given the values
for the phonon and roton parameters which he had adjusted to fit
specific heat data, Landau calculates an approximate value for
$T_{\lambda}$ (2.3~K) also in agreement with
experiment.\cite{note-lambda} He finally explains the
thermomechanical effects -- the fountain effect and the reverse
phenomenon -- and he predicts that heat should propagate as ``second
sound'' instead of diffusing as in classical fluids.

Landau's theory is a remarkable success, and it is still in use
nowadays. Its main features are common to Tisza's previous version
but there is one major difference. The common features are : the
existence of two independent velocity fields; the temperature
variation of the two fluid densities; the non-dissipative flow of the 
superfluid component (but Tisza could not predict the existence of a 
critical velocity for this); the fact that all the entropy is carried 
by the normal component and the propagation of heat as a wave. When
deriving the equations which describe thermomechanical effects, Landau writes
:
``\textit{The formulae 6.1 and 6.4 were deduced already by H. London 
(Proceedings Royal Society 1939) starting from Tisza's ideas}''.
Let me remark that Landau cites Heinz London,\cite{H.London:1939} 
Fritz London's young
brother, and it is very hard to believe that Landau had not noticed
the work of Fritz London, whom he had met in 1932.  The absence
of reference to Fritz London must be intentional. He had 
perhaps personal reasons for this, but I have tried to understand why
 he never believed in the relevance of Bose Einstein condensation 
in the theory of superfluidity. The above sentence 
also means that Landau knew the existence of the two 
notes published in the Comptes-Rendus by Tisza\cite{Tisza:1938b}
in 1938, which are cited by Heinz
London.\cite{H.London:1939}

The major difference between Landau's theory and Tisza's is in the
nature of the normal component: according to Landau it is made of
``quasiparticles'', a new concept he introduces to quantize the
elementary excitations of quantum fluids.  In contrast, Tisza
thinks in terms of ideal gases and proposes that the normal component
is made of the non-condensed atoms.

Shortly after the war, Peshkov did experiments to discriminate between
the predictions by Landau and by Tisza.\cite{Peshkov:1946} Indeed, in the
limit where $T$ tends to zero and according to Landau, the second
sound velocity $c_{2}$ should tend to $c/\sqrt{3}$ where $c$ is the
velocity of the ordinary sound, while Tisza predicted that $c_{2}$
should tend to zero.  At the low temperature meeting which Allen
organized in Cambridge in 1946, a meeting which was called ``LT0'' by
Russell Donnelly 50 years later, Fritz London was asked to give the
opening talk.\cite{London:1946} He explained that Peshkov's
preliminary results\cite{Peshkov:1946} where not yet done at low enough temperature to
discriminate between Landau and Tisza, but Peshkov's experiments soon
showed that Landau was right.\cite{Peshkov:1948} In fact, Fritz London
was very critical about Landau's theory: ``\textit{an interesting
attempt to quantize hydrodynamics\ldots based on the shaky grounds of
imaginary rotons}''.  London must have been rather upset by Landau's
attitude, in particular by his rough rejection of Tisza's model.  Some
authors consider that the two fluid model has been
\textit{independently} discovered by Tisza and by Landau, but this is
not true as we shall see now.  In 1949, Landau wrote a brief report to
Physical Review\cite{Landau:1949} which contains the following note:

``\textit{I am glad to use this occasion to pay tribute to L. Tisza
for introducing, as early as 1938, the conception of the macroscopical
description of helium II by dividing its density into two parts and
introducing, correspondingly, two velocity fields. This made it
possible for him to predict two kinds of sound waves in helium II.
[Tisza's detailed paper (J. de Phys. et Rad. 1, 165, 350 (1940) was
not available in USSR until 1943 owing to war conditions, and I regret
having missed seeing his previous short letters (Comptes-Rendus 207,
1035 and 1186 (1938)).] However, his entire quantitative theory
(microscopic as well as thermodynamic-hydrodynamic) is in my opinion
entirely incorrect.}''

He thus keeps his very abrupt criticism and partly justifies his
former attitude by saying that he was not aware of the details of
Tisza's two fluid model.  But these two letters to the Comptes-rendus,
which Landau pretends that he ``missed'', are those which H. London
cited as his starting point when he derived the ``formulae 6.1 and
6.4'' (see above)!  Since Landau refers to H. London's formulae, he
had read H. London's paper and, consequently, he knew the existence of
Tisza's letters to the Comptes-Rendus.  Could it be then that he had
not read them because they were written in French?  I inquired about
this possibility from A. Abrikosov who sent me the following answer by
mail, on January 15, 2001:

``\textit{Ç Dear Dr.  Balibar,}

\textit{Landau was very able to languages.  He
knew German, English, French and Danish.  Therefore he could read
Tisza's papers in French, the more so that Lifshitz, whom he often
ordered to read papers, instead of doing that himself, didn't know
French\ldots}

\textit{Sincerely yours. Alex Abrikosov}''

Even if E. Lifshitz had perhaps not read these French papers, Landau
knew their existence and it is hard to believe that he had not read
them.  Furthermore, Kapitza also refers to them in his 1941 article
published just before the one by Landau in the Physical Review.
Kapitza measured the thermomechanical effect which is the inverse of
the fountain effect, namely the temperature difference which appears
when superfluid helium flows in a small slit where the normal
component is blocked.  Kapitza uses Landau's theory which is published
as the next article in the same issue.  In his figure, he shows a fit
with a calculation by Landau.  His article was sent the same day (June
23, 1941) as Landau's, which probably means in the same envelope.
It is clear that Kapitza and Landau had a very close collaboration on
this subject.  I cannot believe that they did not share all
information, or that Landau had not read Kapitza's article which
contained his own calculation.  The reference by Kapitza to the two
French articles by Tisza which Landau had ``missed'' is further
evidence that Landau cannot have ``missed'' them.  Even if Landau's
theory is more rigorous and more correct than Tisza's, I consider that
these two works are not \textit{independent}, and that Tisza has a
priority on the two fluid model.

Landau's absence of reference to Fritz London is a different
issue of greater scientific interest.  At this stage, we have to
realize that Landau's 1941 work never mentions Bose nor Fermi
statistics.  In fact he derives his quantization of
hydrodynamics without making any difference between Bose and Fermi
fluids.  Today, of course, we know that degenerate Fermi liquids such
as liquid $^3$He are highly viscous while degenerate Bose fluids are
superfluid.  It means that there is a mistake or some misunderstanding
somewhere in Landau's article.  Where?

After discussing this issue with Grisha Volovik,\cite{Volovik} I
understand that the weak point occurs when Landau claims that there is a
gap between irrotational states and states where the circulation of 
velocity is non-zero.  Landau does not justify this statement.  As is
now well known, it is the work of Bogoliubov\cite{Bogoliubov} which
showed in 1947 for the first time that in a degenerate Bose gas with
weak interactions, there is BEC and there are no individual
excitations at low energy, only collective modes, that is phonons with
a non-zero velocity.  Bogoliubov showed that if dissipation results
from the emission of elementary excitations, it can only occur beyond
a certain critical velocity, (the sound velocity in this case), and
that the motion of the condensate fraction is non-dissipative and
irrotational below this critical velocity.  In 1951, BEC was generalized
by Penrose\cite{Penrose} as
``off-diagonal long range order'' (ODLRO) in the formalism of the density
matrix.  This approach was further developed by Penrose and Onsager in
1956.\cite{Penrose-Onsager} It allows the condensate fraction to be much smaller than
one (the total mass) and irrotational dissipationless motion to occur
below a certain critical velocity.  One has also realized that in most
macroscopic systems, the emission of quantized vortices is another
mechanism which is responsible for a critical velocity smaller than
Landau's.  In other words, the existence of superfluidity is really
linked to BEC, at least to the Bose statistics and the quantization of
vortices.  One could argue, of course that superfluidity
exists in 2D-Bose fluids, where, strictly speaking, there is no BEC.
But there are long range quantum correlations so that vortices are
quantized, and dissipation cannot occur in practice below a certain 
velocity.  In summary, the superfluidity is certainly linked to the
Bose statistics, contrary to Landau's statement.

As for Fermi liquids, it is in fact the hydrodynamics itself which
breaks down.  As Landau was to realize later\cite{Landau:1956}, the
excitations of a degenerate Fermi liquid are Fermi quasiparticles
which travel ballistically over a certain distance and which are
responsible for the divergence of the viscosity in the low temperature
limit.  The existence of an energy gap between rotational and
irrotational states in quantum fluids is simply not true in Fermi
liquids.  This takes us back to the already mentioned question: how
can it be that Landau never referred to BEC nor mentioned Bose
statistics in his theory of superfluidity ?

Perhaps Landau could simply not believe that a
theory of quantum ideal gases (BEC) could apply to liquids with strong
interactions between atoms?  This is the spirit of his criticism of
Tisza's approach (there should be collisions between excited atoms and
condensed atoms).  Furthermore, as would show up later from
Bogoliubov's work\cite{Bogoliubov}, it is true that an ideal Bose gas
with no interactions at all would have a sound velocity equal to zero,
consequently a zero critical velocity: it would not be superfluid!
Eventually, we now know that there is no continuous path from a low
density helium gas to a higher density helium liquid: it has been
predicted\cite{spinodal} and experimentally verified\cite{Caupin}
that there is a range of densities for which helium is unstable,
between two spinodal lines which respectively limit the range of
possible metastability of either liquid or gaseous helium.  For all
these reasons, the most likely interpretation of Landau's absence of
reference to BEC is just that he could not consider that a theory of
quantum gases could apply to a liquid.

However the absence of reference to the Bose statistics needs a
further explanation.  In his 1992 article\cite{Pitaevskii} on
Landau's theory of superfluidity, Lev Pitaevskii writes that
``\textit{Landau was only one step from the very interesting subject
of macroscopic quantum phenomena.  But he never made this step.  And
there is no sense now to guess why\ldots}'' Coming back to this issue
with Landau, Lev Pitaevskii proposed to me another idea which is the
following.  Kapitza and most probably Landau as well considered
superfluidity as a phenomenon analogous to superconductivity.  This
was long before the BCS theory and of course superconductivity occurs
in a Fermi system of electrons.  Since the same phenomenon occurred in
both quantum fluids (Bose and Fermi), Landau could perhaps not admit
that superfluidity was related to the quantum statistics.  Whatever
Landau really thought, a possible comparison of $^3$He and \He4
progressively appeared as a crucial test.  In his book, London
insisted on the importance of such a test.\cite{London:book}  As soon
as $^3$He was available in large enough quantities, a test was made of the
possible superfluidity of $^3$He, which was found to be non-superfluid down
to 1~K, in strong support to London's and Tisza's approach.  This
experiment was done by D.W. Osborne, B. Weinstock and B.M.
Abraham\cite{Osborne} in 1949.

As an aside, let me
mention that B. Abraham had joined the Manhattan project during the
war and owned a patent for the Lithium-Tritium compound to be used
in H-bombs.  Let me  mention further that Landau also participated to the
building of the H-bomb, but the Soviet one of course, and despite the
severe conflict which opposed Kapitza and Beria in this enterprise.
Beria forced Kapitza to leave his scientific position and activity at
the Institute for Physical Problems because of their conflict.  Landau
kept working for the bomb, apparently because this was a way for him
to be protected against any further problems with the Soviet
regime.\cite{Gorelik} Later, Beria was assassinated and Kapitza
recovered his position at the Institute for Physical Problems. When
Stalin died, Landau left the Soviet H-bomb program.\cite{Gorelik}

Coming back to superconductivity and the superfluidity of $^3$He, we
know that the BCS theory considers the condensation of Cooper pairs
which obey the Bose statistics, and that superfluidity was also
discovered in liquid $^3$He at a temperature low enough (about 2.5~mK)
that $^3$He atoms could form pairs.\cite{Osheroff,Leggett1}

As for rotons, their existence was proven by inelastic neutron
scattering experiments.\cite{Henshaw}  It also happens that, for my
PhD work, I studied quantum evaporation and obtained the first
experimental evidence that, at low enough temperature, a heat pulse
decomposes into ballistic phonons and rotons, and that individual
rotons can evaporate atoms with a minimum kinetic energy of
1.5~K.\cite{Balibar:1978}  This phenomenon had been predicted by P.W.
Anderson as an analogue of the photoelectric effect.\cite{Anderson}
A.F.G. Wyatt and his group have performed a long quantitative study of
it.\cite{Hope,Brown,Tucker} Today, there is no doubt that rotons exist, only
controversies remain on their physical picture.  Landau had first proposed that
they were vortices of atomic size and later considered them as part of 
the phonon spectrum. Surprisingly, Feynman insisted on Landau's first picture
by considering that a roton could be an elementary vortex loop.\cite{Feynman}  In my
opinion, rotons are phonons with a wavelength equal to the interatomic
distance. Their low energy is a signature of the local order 
which had already been mentioned by London.  As expressed
by Nozi\`eres, rotons are ``ghosts of a Bragg peak''\cite{Nozieres}
(in fact, this idea was already present in the work of other
authors\cite{Horner,Pomeau}). 
This is because Feynman showed that, under certain approximations, the dispersion
relation $\omega(q)$ for elementary excitations is related to the
static structure factor $S(q)$ of liquid helium by the simple relation

\beq
\hbar\omega (q)  = \frac{\hbar^2 q^2}{2 m S(q)}\: .
\label{eq:feynmann}
\eeq

As explained to me by G. Volovik\cite{Volovik}, this relation only
requires that the wavefunction describing the fluid is symmetric as it
has to be for a Bose fluid.  The above equation shows that, if there
is some short range order in this liquid, that is a large probability
to find an atom at a distance which is the average interatomic
distance from another atom, in other words a large peak in the
structure factor $S(q)$, then there has to be a roton minimum in the
relation $\omega (q)$.  One should not associate superfluidity with
the existence of a roton minimum; Landau introduced rotons to
calculate the specific heat of liquid helium and then explained that
their existence limits the maximum value of the critical velocity.  In
reality rotons are precursors of solidification, and their existence
works against superfluid order.  In the superfluid gases which have
been discovered in 1995,\cite{Cornell,Ketterle} there is
superfluidity and no rotons because the system has weak interactions.
In the superfluid liquid, an instability is predicted to occur if the
roton minimum goes to zero -- if rotons become soft -- in which case
the dispersion relation resembles that of a crystal in the extended
zone scheme.  The existence of such an instability is under present
investigation in my research group.\cite{Ishiguro} Above the lambda
temperature, rotons still exist, they are no longer well defined modes
with a long lifetime but this is also true for the rest of the
dispersion curve.

Landau was right in a sense (rotons exist)
but wrong concerning his first interpretation or physical picture (they are not
elementary vortices, nor essential for superfluidity).

\section{Later developments}

In my opinion, London and Tisza had found part of the truth and Landau
had found a complementary part of the truth.  Unfortunately, neither
London nor Landau lived long enough to realize that a full theory
should combine their respective approaches.  Fritz London died of a
heart attack in 1954.  Landau was severely injured in a car accident
shortly before receiving his 1962 Nobel prize. The car accident
occurred on January 7, 1960, he was in coma for a long time and
suffered so much afterwards that he could never work anymore till he
died in 1968.  Of course he could not go to Stockholm and receive his
Nobel prize in person. In my
opinion, he would probably have shared this prize with London if
London had not died before.  London had been proposed for the Nobel
prize by Einstein.  A few years before arriving to the famous
BCS theory with Leon Cooper and Robert Schrieffer, John Bardeen 
also recognized the great importance
of London's work on superconductivity (the introduction of a
macroscopic wave function) as the basis of his work on
the same subject;
in a letter sent to London on
December 9, 1950, he had written : 

``\textit{Dear Prof. London}

\textit{You may be interested in the enclosed manuscripts  on
superconductivity; they are both based on your approach}''.\cite{Meyer}

Bardeen's admiration for London's work must be the 
reason why, when he received his second Nobel prize in 1972
(he shared this one with Cooper and Schrieffer for the ``BCS'' theory
of superconductivity but he had already shared one with Schockley and
Brattain in 1956 for the discovery of the transistor), 
he decided to donate his part of the Nobel Prize to Duke University.
 The purpose was to create an endowment to enable funding a yearly
lecture at Duke University in the honor of Fritz London and also to
finance the Fritz London Prize for distinguished work in Low
Temperature Physics.  This Prize, which has become very prestigious,
was given for the first time to N. Kuerti in 1958 for his work on nuclear magnetism.  I was surprised
to see that the second London Prize was given to Landau in 1960 (the
third one was given to John Bardeen in 1962).  Of course, Landau's
exceptional achievements in physics deserved more than the London
prize, but it means that the London prize jury totally ignored the
controversies and personal conflicts which opposed London and
Landau.  In an e-mail which he sent me on January 21, 2001, Tisza
wrote:

``\textit{I know that Landau had no high regard for London.  I think he
was wrong and hurt his own science for yielding to his spite.  London
disliked Landau, and I did what I could to temper his feelings when
writing his ``Superfluids''.  I suspect that they had an unpleasant
interaction in 1932 when Landau traveled in the West, but this will
remain an unsolved mystery.}''
    
I am pleased to see that science is sometimes more important than
personal conflicts.  On June 17, 2005, I received another message from
Laszlo Tisza where he commented on the London prize:

``\textit{Dear Sebastien,\\
\ldots
Yesterday I was leafing through old correspondence and I found a letter
in which I nominated Landau for the Prize. I am sure I was not alone. I 
was actually at LT-7 in Toronto when the Prize was announced. It is actually
unconscionable of Landau not to have taken note of the remarkable Simon - London
work on helium in Oxford 1934-35! I never heard a word of it while at UFTI. 
All he said was that London was not a good physicist.
I am looking forward to your book to straighten out matters.
With warmest regards,
Laszlo}''

Laszlo Tisza himself supported the nomination of Landau for the London
prize!  He had recognized the superiority of Landau's
two fluid model on his early theory and he did not want to
be upset by any personal criticism which he considered as secondary.  Fifty
years later, he still thinks the same way.

Kapitza was awarded the Nobel prize in 1978.  This was 16 years after
Landau and 41 years after he had sent his historical letter to
\textit{Nature}.  In his speech, he noticed this surprising delay and
talked about a different subject (nuclear fusion).  I do not know if
the Nobel prize jury ever considered the possibility of dividing a Nobel prize on
superfluidity between Kapitza and Allen.  Perhaps some physicists
considered that Kapitza had some priority on Allen and it was
difficult to find agreement.  I have already detailed my opinion about
this issue.

I wish to conclude with another quotation from Tisza. At the end of
his talk for the hundredth anniversary of the Hungarian physical
society in 1991\cite{Tisza:1991}, he wrote:

``\textit{If history has a lesson, it is that the ``winner takes all''
attitude deprives one of the pleasure of being the heir to the best of
different traditions, even while avoiding their intolerance against
each other}.''

Tisza was squeezed between London and Landau whose approaches of the
theory of superfluidity were rather different.  In fact London
considered the ground state of liquid helium and Landau its excited
states.  It took quite a long time to unify their respective points of
view, even after Bogoliubov's work.\cite{Bogoliubov}  As for Tisza,
some of his theory was wrong but he had introduced many of the
fundamental ideas which were later developed by Landau.  Furthermore,
Landau's theory was not entirely correct either.  In conclusion, one
should certainly attribute the discovery of the theory of
superfluidity not only to London and Landau, but also to Tisza.

\section{Acknowledgments}
I wish to congratulate Frank Pobell for his remarkable achievements in
low temperature physics and for letting me take the opportunity of
this special issue of the Journal of Low Temperature Physics to recall
the long and rich history of the discovery of superfluidity.  I am
very grateful to many colleagues for very fruitful discussions and
many suggestions after careful reading of preliminary versions of this
article, especially to Laszlo Tisza, Lev Pitaevskii, Grisha Volovik,
Horst Meyer, Allan Griffin, David Edwards and Roger Bowley.


\begin{thebibliography}{9}
\bibitem{note-lambda}However, note that when writing that ``{\it the temperature at which
$\rho_{n}/\rho$ equals unity is the $\lambda$-point of helium}'', Landau
does not necessarily mean that there is a real phase transition at
this point, it could be just a crossover from a quantum to a classical
behavior.
\end{thebibliography}
\end{document}